 \documentclass[preprint2]{aastex}

\def\dsm{$M_\odot$}
\def\teff{$T_{\rm eff}$}
\def\dov{$\delta_{\rm ov}$}

\shorttitle{The effects of the overshooting on main-sequence turnoffs of star clusters}
\shortauthors{Wuming Yang}

\begin{document}


\title{The effects of the overshooting of the convective core
on main-sequence turnoffs of young- and intermediate-age star clusters}
\author{Wuming Yang\altaffilmark{1}, Zhijia Tian\altaffilmark{2}}
\affil{$^{1}$Department of Astronomy, Beijing Normal University,Beijing 100875, China}
\email{yangwuming@bnu.edu.cn}
\affil{$^{2}$Department of Astronomy, Peking University,Beijing 100871, China}

\begin{abstract}
Recent investigations have shown that the extended main-sequence turnoffs (eMSTOs)
are a common feature of intermediate-age star clusters in the Magellanic Clouds.
The eMSTOs are also found in the color-magnitude diagram (CMD) of young-age star
clusters. The origin of the eMSTOs is still an open question. Moreover,
asteroseismology shows that the value of the overshooting parameter
$\delta_{\rm ov}$ of the convective core is not fixed for the stars
with an approximatelly equal mass. Thus the MSTO of star clusters
may be affected by the overshooting of the convective core (OVCC).
We calculated the effects of the OVCC with different $\delta_{\rm ov}$
on the MSTO of young- and intermediate-age star clusters. \textbf{If
$\delta_{\rm ov}$ varies between stars in a cluster,}
the observed eMSTOs of young- and intermediate-age star clusters
can be explained well by the effects. The equivalent age spreads of MSTO
caused by the OVCC are related to the age of star clusters and are in good
agreement with observed results of many clusters. Moreover,
the observed eMSTOs of NGC 1856 are reproduced by the coeval
populations with different $\delta_{\rm ov}$. The eMSTOs of star clusters
may be relevant to the effects of the OVCC. The effects of the OVCC
\textbf{are similar to that of rotation in some respects. But the effects
cannot result in a significant split of main sequence of young star clusters
at $m_{U}\lesssim 21$.} The presence of a rapid rotation can make
the split of main sequence of young star clusters more significant.
\end{abstract}


\section{INTRODUCTION}

The double or extended main-sequence turnoffs (eMSTOs) were discovered
in the color-magnitude diagram (CMD) of intermediate-age massive star clusters
in the Magellanic Clouds \citep{mack07, mack08, glat08,
milo09, gira09, goud09, goud11, kell12, piat13}. One interpretation
of the eMSTOs is that the clusters have experienced an extended
star-formation histories (eSFH) with a duration of $\sim 100-700$ Myr
\citep{mack08, glat08, milo09, gira09, rube10, goud11, goud14, corr14},
which disagrees with classical understanding of star clusters being
simple stellar populations (SSPs).

The eMSTOs were also discovered in young clusters NGC 1856 \citep{corr15,
milo15}, NGC 1755 \citep{milo16a}, NGC 1850 \citep{bast16}, and NGC 1866
\citep{milo16b}. Moreover, NGC 1856, NGC 1755, and NGC 1866 are found to
exhibit dual main sequences (MS) below their MSTO \citep{milo15, milo16a,
milo16b}.

An alternative interpretation of the eMSTOs is the effects of star rotation
\citep{bast09, yang13, bran15, dant15, nied15a}.
\cite{yang13} show that the extension of MSTO caused by star rotations
is related to the rotation rate of stars, the efficiency of rotational mixing,
and the age of star clusters. A relatively high rotation rate and a high
efficient rotational mixing are required to explain the eMSTOs of young
clusters [see Figure 8 in \citet[but see Niederhofer et al. 2015a and D'Antona
et al. 2015]{yang13}]. \cite{nied15a} claimed that the eMSTO of NGC 1856 can be
explained by a rotation of $0.5$ times the Keplerian rotation rate
($\Omega_{\mathrm{cr}}$). But in order to explain the dual MSs of clusters
NGC 1856 and NGC 1755, the rotation rate of $0.9$ $\Omega_{\mathrm{cr}}$
is required \citep{dant15, milo16a}. \textbf{A large number of rapid rotating
stars have been found in NGC 1850 and NGC 1856 by \cite{bast17}. }

\textbf{However, neither stellar models with different ages only, nor rapid
rotating models with different rotation rates, properly reproduce the observed
split MS and eMSTO of NGC 1866 \citep{milo16b}. The populations with both different
ages and different rotation rates are needed to explain NGC 1866 \citep{milo16b}.}
Moreover, \cite{dant15} stated that their rotating models fail to reproduce
the stars ``after the end of the central H-burning phase" of NGC 1856.
\textbf{However, these stars might be stars with decretion
disks \citep{bast17} seen nearly edge on, so they suffer from extinction
which pushes them into this region.}

Another coeval interpretation of the eMSTOs is interacting binaries
\citep{yang11, liz12, liz15, liz16}. \cite{yang11} showed that interacting
binaries including merged binary systems and the binaries with mass transfer
can lead to both the eMSTOs and the dual red-clumps. The effects of the
interacting binaries on the CMDs of some clusters should not be neglected,
although the number of the interacting binaries in a cluster could be not
enough to explain the eMSTOs alone.

One of the important predictions of the eSFH scenario is that the ongoing
star-formation should be observed in young star clusters with an age of a few
hundred Myr. However, up to now, the expected ongoing star-formation is not
observed in young clusters with age beyond 10 Myr \citep{bast13, bast16, cabr14,
cabr16, nied15b}. \cite{goud11, goud14} and \cite{corr14} argued that the
eMSTOs can occur only in clusters with masses larger than a threshold of
about $10^{5}$ \dsm{} and with escape velocity greater than $12-15$ km s$^{-1}$.
However, the eMSTOs of NGC 1755 \citep{milo16a} and NGC 411 \citep{lic16a} would
represent a challenge for this scenario. \textbf{Furthermore, the observation
that there exists a strong correlation between cluster age and the inferred
age spread as found by \cite{nied15a} also rules out an actual age spread being
the origin of the eMSTO.}

\cite{lic14} analyzed the sub-giant branch (SGB) of NGC 1651
harbouring an eMSTO and found that the SGB is narrower and offsets from
what would be inferred from the eMSTO region if the large age spreads would
be present within the cluster. Similar results were found in NGC 1806 and
NGC 1846 \citep{bast15} and NGC 411 \citep{lic16a}. Hence, they concluded
that age spreads are not likely to be the cause of the eMSTO phenomenon.
However, \cite{goud15} found that the cross-SGB profiles of NGC 1651, NGC 1806,
and NGC 1846 are consistent with their cross-MSTO profiles when the latter
are interpreted as age distributions. Conversely, their SGB morphologies
are inconsistent with those of simulated SSPs.
The origin of the eMSTOs is still an open question.

The overshooting of the convective core (OVCC) can bring more
hydrogen-rich material into H-burning core, which significantly
prolongs the lifetime of the burning of core hydrogen and enhances
the He-core mass left behind. The distance of the overshooting of
a convection is defined as \dov{}$H_{p}$, where \dov{} is a free
parameter and $H_{p}$ is the local pressure scale-height.
Recently, \cite{yang16b} developed a method to determine the size of the convective core
including the overshooting region from observed oscillation frequencies of
low-degree $p$-modes. It was found that the value of \dov{} is variable
for stars with an approximatelly equal mass. For example, the value
of \dov{} is $0.2$ for KIC 9812850 with $M\simeq1.48$ \dsm{} \citep{yang16b},
$\sim1.4$ for KIC 2837475 with $M\simeq 1.50$ \dsm{} \citep{yang15},
$0.9-1.5$ for Procyon with $M\simeq 1.50$ \dsm{} \citep{guen14, bond15},
$\sim0.6$ for HD 49933 with $M\simeq 1.28$ \dsm{} \citep{liu14}, and
$1.7-1.8$ for KIC 11081729 with $M\simeq 1.27$ \dsm{} \citep{yang15b}.
\textbf{The typical errors of the value of \dov{} are $0.1-0.2$.}
If a variable overshooting exists in stars with masses larger than $1.50$ \dsm{},
the MSTO of young- and intermediate-age star clusters would be affected
by the overshooting.

In this work, we mainly focus on whether the eMSTOs of young- and
intermediate-age star clusters can be explained by the effects of the OVCC.
The paper is organized as follows: we show our calculation results in Section 2,
and the results are compared with observations in Section 3,
then we discuss and summarize the results in Section 4.

\section{CALCULATION RESULTS}
\subsection{Evolutionary tracks}
In order to study the effects of overshooting of the convective core on the MSTO
of star clusters, we computed a grid of evolutionary models with the initial
metallicity $Z_{i}=0.008$, \dov{} in the range of $0.0-0.6$ with a resolution
of $0.2$, supplemented by $0.7$, and masses between $0.9$ and $6.0$ \dsm{}.
The resolution of mass varies from $0.005$ to $0.03$ \dsm{}. We used
the Yale Rotation Evolution Code (YREC) \citep{pins89, yang07, yang16a}
to construct the models in its non-rotation configuration.
The OPAL equation-of-state table EOS2005 \citep{roge02}
and OPAL opacity table GN93 \citep{igle96} were adopted, supplemented by
the \cite{alex94} opacity tables at low temperature. The diffusion and settling
of both helium and heavy elements are computed by using the diffusion coefficients
of \cite{thou94} for models with a mass less than $1.3$ \dsm{}. Convection is treated
according to the standard mixing-length theory. The value of the mixing-length
parameter is fixed at $1.75$. The full mixing of material is assumed
in the overshooting region. All models are evolved from zero-age MS
to the red giant branch (RGB).

\begin{figure}
\centering
\includegraphics[scale=0.6, angle=-90]{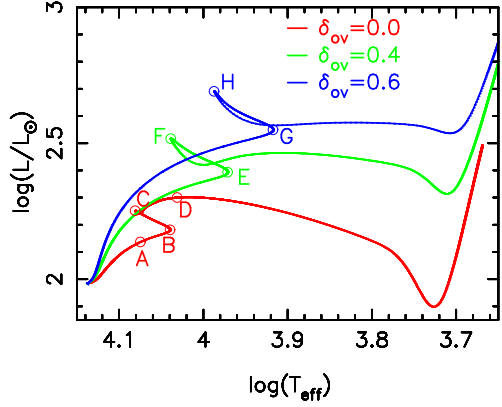}
\caption{Hertzsprung-Russell diagram of models with $M = 3.0$ \dsm{} with
different \dov{}. The values of the central hydrogen $X_{c}$ at points A, B,
C, and D are about $0.27$, $0.05$, $0.0005$, and $0$, respectively.
The value of $X_{c}$ is about $0.025$ at point E, $\sim8\times10^{-6}$
at point F, $\sim0.016$ at point G, and $\sim8\times10^{-12}$ at point H.}
\label{fig1}
\end{figure}

Figure \ref{fig1} shows the evolutionary tracks of models with
$M=3.0$ \dsm{} but with different \dov{}. The evolutionary
tracks are obviously affected by the OVCC. The model with a large
\dov{} mainly exhibits to be bluer than the model with a smaller
\dov{} at a given age and the lifetime of the burning of core hydrogen
is significantly prolonged, but the lifetime of SGB is shrunk.
For example, the lifetime of the SGB of the model with $M=3.0$ \dsm{}
and \dov{} = 0 is $12.5$ Myr, which is $4.5\%$ of its MS lifetime.
But the lifetime of the SGB of the model with $M=3.0$ \dsm{}
and \dov{} = $0.4$ is only $0.7$ Myr, which is about
$0.2\%$ of its MS lifetime. The Hertzsprung gap of the model
with a large \dov{} is obviously narrower than that of
the model without the OVCC.

The values of the central hydrogen $X_{c}$ at points B, E,
and G in Figure \ref{fig1} are about $0.05$, $0.025$, and $0.016$,
respectively, while the values of $X_{c}$ at points C, F, and H
are around $0.0005$, $8\times10^{-6}$, and $8\times10^{-12}$, respectively.
Compared with the whole MS lifetime of stars with a large \dov{},
the timescale of their MS hook is very short. Thus we define
the points B, E, and G as the end of central H-burning phase.
Due to the fact that more hydrogen-rich material is brought into
H-burning core by the OVCC, the stars with the OVCC produce more
nuclear energy. Part of the energy makes the stars more luminous.
The other is transformed into internal energy,
which leads to the fact that the stars expand obviously,
i.e., the radius of the stars increases significantly.
As a consequence, the effective temperature of the stars decreases
at the end of central H-burning phase. And the star with a larger
\dov{} is brighter and redder than the star with a small \dov{}
at the end of central H-burning phase. Thus, the MS width and MSTO
of a star cluster must be extended by the presence of the OVCC with
different \dov{}.

Moreover, due to the fact that the timescale of the MS hook
of the model with a large \dov{} is significantly shrunk,
the structure of MS hook of a young star cluster including
many stars with the large \dov{} is hard to be exhibited in
the observed CMD of the cluster.

\subsection{Isochrones}
\begin{figure}
\centering
\includegraphics[scale=0.36, angle=-90]{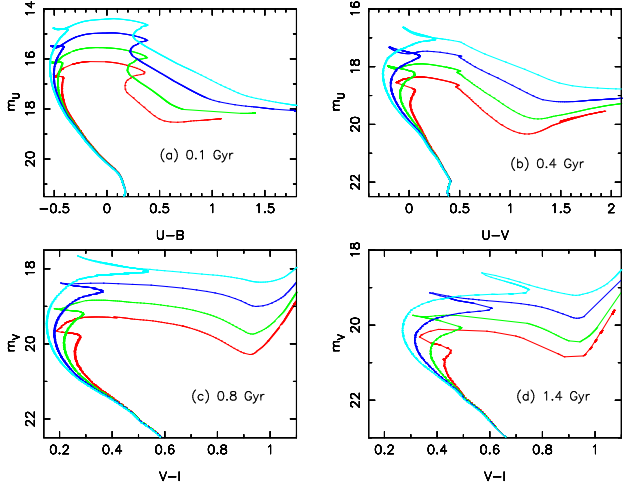}
\caption{The CMDs of isochrones of models with different \dov{}.
Panels $a$, $b$, $c$, and $d$ show the isochrones with the
age of $0.1$, $0.4$, $0.8$, and $1.4$ Gyr, respectively. The red
lines represent the isochrones of models with \dov{}$=0$. The
green, blue, and cyan lines show the isochrones of models with
\dov{}$=0.2, 0.4$, and $0.6$, respectively.}
\label{fig2}
\end{figure}

\begin{figure}
\centering
\includegraphics[scale=0.6, angle=-90]{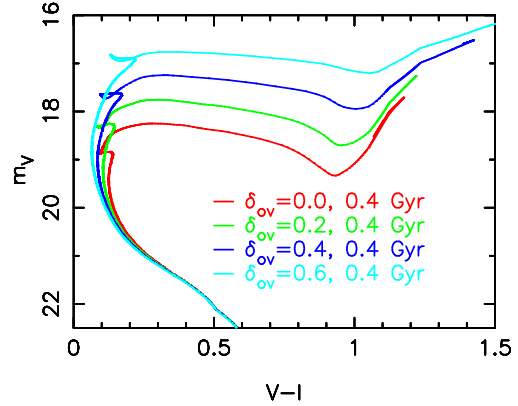}
\caption{The CMD of isochrones with different \dov{} but with the
same age.}
\label{fig3}
\end{figure}

In order to understand the effects of an OVCC on the MSTO of
young- and intermediate-age star clusters, we calculated
isochrones with different \dov{}. To obtain the CMDs of the isochrones,
the theoretical properties ([Fe/H], \teff{}, $\log g$, $\log L$)
of models are transformed into colors and magnitudes using the color
transformation tables of \cite{leje98}. A distance modulus of $18.6$
is adopted, and the value of $0.12$ is adopted for $E(B-V)$ in the
calculation. Figure \ref{fig2} shows the CMDs of isochrones of models
with different \dov{}. Comparing the isochrones of models
with a large \dov{} to those of models with a smaller \dov{}, one can find
that the upper MS and the MSTO of young- and intermediate-age star clusters
are extended by the OVCC when the different OVCCs exist in clusters.
However, the lower MS of the clusters are almost not affected. For the isochrones
with the age of $400$ Myr, the separation between the isochrone of models with \dov{}
$=0.4$ and that of models with \dov{} $=0.0$ is about 0.1 mag in color U-V at
m$_{U}=19.5$, but it is only around $0.02$ mag at m$_{U}=21$. However,
the separation is about $0.03$ mag in color V-I at m$_{V}=19.5$.
The eMSTO of young clusters is better visible in U-V than in V-I
(see the panel $b$ of Figure \ref{fig2} and Figure \ref{fig3}).
These characteristics are similar to those found in NGC 1856
and NGC 1755 \citep{milo15, milo16a} \textbf{and are also expected
in the rotational scenario.}

\section{COMPARISON WITH OBSERVATIONAL RESULTS}
\subsection{The equivalent age spread caused by the OVCC}

\begin{figure}
\centering
\includegraphics[scale=0.6, angle=-90]{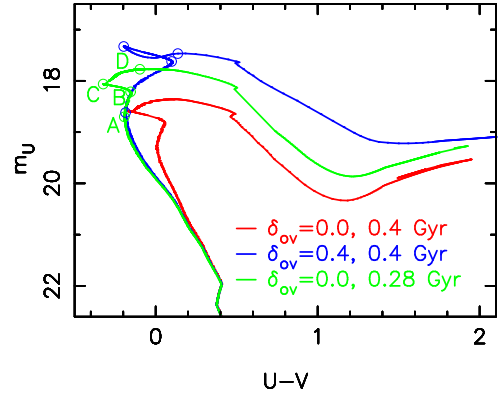}
\caption{An example of determination of equivalent age spread.
The values of $X_{c}$ of models at points A, B, C, and D
on the green isochrone are about $0.27$, $0.05$, $0.0005$,
and $0$, respectively.}
\label{fig4}
\end{figure}

We assumed that a distribution of \dov{} such as a bimodal of
$0$ and $0.4$ exists in a star cluster. The isochrones of models
with \dov{} $=0$ are chosen as the standard one.
The isochrones of models with a large \dov{} are compared to
those of models with \dov{} $=0$ to obtain the equivalent age spread
caused by the OVCC. Hereafter, the beginning and the end points
of MS hook of an isochrone are labeled as B and C, respectively.
The local bluest point (local minimum U-V) of MS is marked
as point A, and the brightest point (minimum $m_{U}$) after
point C is labeled as point D (see Figure \ref{fig4}).
The values of the central hydrogen $X_{c}$
of models at points A, B, C, and D on the isochrone with age $=280$
Myr are about $0.27$, $0.05$, $0.0005$, and $0$, respectively.
According to the initial mass function, the more massive the stars,
the less the number of the stars. Moreover, the more massive the stars,
the faster their evolutions. The stars between point C
and D ($X_{c}$ between about $0.0005$ and 0 for \dov{} $=0$
but between about $8\times10^{-6}$ and 0 for \dov{} $=0.4$) should be
rare in a young star cluster. Thus we compare the non-overshooting
isochrone between point A and B to the isochrone with an overshooting
to obtain the equivalent age spread. For example, inspected by eyes,
the non-overshooting isochrone between point A and B in Figure \ref{fig4}
is considered to overlap with the overshooting isochrone with \dov{} $=0.4$
and age $=400$ Myr. Thus, the equivalent age spread caused by the
overshooting with \dov{} $=0.4$ is $120$ Myr when the age of the
cluster is $400$ Myr. The equivalent age spreads for other isochrones
with different \dov{} and ages are obtained in the similar way.
The  ``points C and D'' on the isochrone with age $=400$ Myr and \dov{} $=0.4$
cannot be matched by those of any isochrone with \dov{} $=0$. Thus,
we cannot obtain the equivalent age spread through comparing
the points C or D of isochrones with different \dov{}.

Due to the fact that the models with a large \dov{} are obviously
brighter and redder than the models with a smaller \dov{}
at the end of central H-burning phase, the end point of
the central H-burning phase of an isochrone with \dov{}$=0.4$
cannot be reproduced by that of the isochrones with \dov{}$=0$
(see Figure \ref{fig4}). Thus, when one uses the isochrones with
a small \dov{} to fit the CMD of a cluster including many
stars with a larger \dov{}, one could find that there are
many stars that cannot be reproduced after the end of
the central H-burning phase.

In order to understand whether the effects of the OVCC can explain
the observed eMSTOs of star clusters, the equivalent age spreads
caused by the OVCC are extracted as mentioned above and compared
to observed ones. \textbf{We do a similar experiment as \cite{nied15a}.}
Figure \ref{fig5} shows the comparison between
the equivalent age spreads and those observed by \cite{goud14},
\cite{nied15b}, \cite{corr14, corr15}, \cite{milo15, milo16a},
and \cite{bast16}. The values of the age spreads of Goudfrooij
sample adopted here are the values of $W20_{\mathrm{MSTO}}$,
which are considered to cannot be explained by rotating models
\citep[but see \textbf{Brandt \& Huang 2015 and}
Niederhofer et al. 2015a]{goud14}. The equivalent age spreads
increase with an increase in age of star clusters and \dov{}, and are in
good agreement with the observed ones, which shows that the observed eMSTOs
of young- and intermediate-age star cluster can be interpreted by the effects
of the OVCC. \textbf{The inferred age spreads of a given \dov{} shown
in Figure \ref{fig5} are relative to the same age isochrone with \dov{}
$=0$, i.e. the extension of MSTO caused by a \dov{} at a given age is relative
to the isochrone with the same age but with \dov{} $=0$. }
Keep in mind that a single value of \dov{} cannot produce
any eMSTO in the CMD of a cluster. \textbf{And the trend between
the age of star clusters and the inferred age spread is also expected
in the rotational scenario.}

Figure \ref{fig5} shows that the equivalent age spreads caused by
the OVCC increase with an increase in age of star clusters.
However, when the age of star clusters is larger than 2 Gyr, the equivalent
age spreads caused by \dov{} = 0.2 almost not increase with an increase in age
of clusters. This is because the convective core of low mass stars shrink
during their MS. But for more massive stars, the convective core increases
in size during the initial stage of MS of these stars before the core
begins to shrink. The larger the \dov{}, the later the core begins to shrink.
In the late stage of the MS of stars, when the core shrinks to a certain degree,
the chemical composition of the core including the overshooting region is
mainly helium. At this time, the effects of the overshooting on the evolution
of stars are not significant. The smaller the value of the \dov{}, the earlier
this scenario appears. Therefore, the equivalent age spreads caused by
the OVCC cannot increase continuously with
an increase in age of star clusters. As a consequence, the equivalent
age spreads caused by \dov{} = 0.2 reach a maximum of about $330$ Myr
at the age of about $2.2$ Gyr.

\textbf{\cite{nied16} have shown that observationally there is 
a peak in the inferred age spread at age $\sim 1.7$ Gyr, which 
is consistent with that expected by rotation models
of \cite{yang13}. The equivalent age spread is still 
present at age $>2$ Gyr in OVCC scenario. But the width 
of eMSTO caused by the OVCC reaches a maximum in color V-I
of CMD of star clusters at age $\sim 1.4$ Gyr, and then decreases 
with an increase or decrease in age (see panels $c$ and $d$ of
Figure \ref{fig2} and Figure \ref{fig6}). The equivalent age 
spread caused by \dov{} $= 0.2$ is about 300 Myr at age $=3.2$ Gyr.
But the separation between the MS of models with \dov{} = 0.2 
and that of models with \dov{} = 0 is only $0.026$ mag 
at $m_{V}=21.9$ mag, which is comparable with the observational
error. The separation is around $0.044$ mag at $m_{V}=21.6$ mag
(see panel $c$ of Figure \ref{fig6}). When the age of clusters 
is about $3.7$ Gyr, the MS of models with \dov{} $=0$ is very 
close to that of models with \dov{} $=0.4$ in the CMD; 
the separation between the \dov{} $=0.4$ isochrone and 
the \dov{} $=0.2$ isochrone is about $0.04$ mag at $m_{V}=21.8$ mag. 
The maximum separation between the MS of the \dov{} $=0.2$ or $0.4$ 
isochrone and that of the \dov{} $=0$ isochrone in color V-I at age $=1.4$
Gyr is about $1.3$ times as large as that at age $=0.8$ Gyr, and is 
around $2$ times as large as that at age $=2.5$ Gyr. The eMSTO caused 
by a large \dov{} should be present in the CMD of star clusters 
with ages $> 2.0$ Gyr unless the fraction of stars with the large
\dov{} is too small. But it is not as evident as that in 
intermediate-age star clusters. This trend is not reflected by the 
equivalent age spreads. }

\subsection{Comparison with NGC 1856}
\subsubsection{The eSFH scenario}

NGC 1856 exhibits the extended MSTOs. \cite{nied15a} claimed
that the eMSTOs of NGC 1856 can be explained by the effects of rotation with
initial rotation rate $\Omega_{i}=0.5$ $\Omega_{cr}$. 

In Figure \ref{fig7}, we compare the data of NGC 1856 to models with different
\dov{}, which shows that the stars at the end of the central H-burning phase
and the subgiants can be reproduced by the models with \dov{} $=0.4$ and
$0.6$ but cannot be reproduced by the models with \dov{} $=0$ or $0.2$.
The age of a star cluster obtained from isochrones
increases with an increase in \dov{}. For example,
when one uses the isochrones of models with \dov{} $=0$ to fit the
CMD of NGC 1856, one could obtain an age of $\sim300$ Myr;
however, when one uses the isochrones of models with \dov{} $=0.4$ to fit
the CMD, an age of $\sim500$ Myr would be obtained. Moreover, 
Figure \ref{fig7} also shows that one would obtain a smaller age spread 
if one uses the isochrones with a small \dov{} to fit the eMSTOs of NGC 1856.
Panel $c$ of Figure \ref{fig7} shows that the eMSTOs of NGC 1856
can be explained by an age spread no more than $300$ Myr of models
with \dov{} $=0.4$. 

\textbf{
NGC 1856 also exhibits the dual MSs. The dual MSs of NGC 1856
are separated by $\sim0.1$ mag in color U-V at $m_{U}$ $\sim20.0$ and merge
together $\sim1.5$ magnitudes below \citep{milo15}. In order to explain
the split MS, the rotation as high as 0.9 $\Omega_{cr}$ is required \citep{dant15}.
Moreover, the simulated populations of \cite{dant15} cannot reproduce
the stars near the end of the central H-burning phase (see their figure 4).
But this might be due to stars with decretion disks seen nearly 
edge on \citep{bast17}.
}

The separation between the isochrone with age $=700$ Myr and that 
with age $=400$ Myr is $\sim0.1$ mag in color U-V at $m_{U}$ $\sim20.0$,
which is in good agreement with observation of \cite{milo15}, 
and is only $\sim0.02$ mag at $m_{U}$ $\sim21.5$. \textbf{Recently,}
\cite{lic16c} analyzed the data of NGC 1856 and speculated that 
the rapid stellar rotation scenario is the favored explanation 
of the multiple sequences of NGC 1856 rather than eSFH scenario.

\subsubsection{Rapid rotation scenario}
We computed
rapid rotation models with $\Omega_{i}=0.5$ $\Omega_{cr}$ and with the
rotational mixing efficiency for the Sun \citep{pins89, yang07}. In
Figure \ref{fig8}, we compare the evolutionary tracks of our models
with those of Geneva models \citep{geor13}, which shows that our rotating
and non-rotating models with \dov{} $=0$ are comparable with Geneva models.
Figure \ref{fig8} shows that the effects of the OVCC with \dov{} $=0.4$
on the end point of central H-burning phase cannot be mimicked by
the effects of the rapid rotation. When both the OVCC and
the rapid rotation exist in a star, the effects of the OVCC will
play a dominant role because the OVCC can mix all material in the
overshooting region with that in the convective core but the rotational
mixing only brings a small part of material in the shell on the top of
the overshooting region into the convective core.

Due to the centrifugal effect, rotating models are fainter and redder
than non-rotating overshooting models in the early stage of MS.
However, in the late stage of MS, due to the fact that slightly
more H-rich material is brought into H-burning core by rotational
mixing, rotating models produce slightly more nuclear energy.
Thus, they exhibit slightly brighter than non-rotating
overshooting models (see the panel $b$ of Figure \ref{fig8}).
However, the rotational mixing in the radiative region also leads to
an increase in the mean density of the region, i.e. a decrease
in radius of stars, which results in the fact that rotating models
have a higher effective temperature at the end of central H-burning
phase. As a consequence, rotating overshooting models are slightly
brighter and bluer than non-rotating overshooting models at the end
of the central H-burning phase.

In panels $a$, $b$, and $c$ of Figure \ref{fig9}, we compare rotating and
non-rotating models with different \dov{} to the data of NGC 1856,
which show that the effects of the rotation of $0.5$ $\Omega_{cr}$ are not
enough to explain the eMSTOs of NGC 1856 in our calculations.
Due to the fact that the centrifugal effect plays a dominant role
in the early stage of MS phase of stars whose envelope is radiative
(with masses $\gtrsim 1.3$ \dsm{}) and leads to the fact that rotating models
are fainter and redder than non-rotating models, the MS of
rotating models is almost parallel with that of non-rotating
models between $m_{U}$ $\sim19.5$ and $m_{U}$ $\sim22$.
The separation between the MS of rotating models and that
of non-rotating models is about $\mathbf{0.016}$ mag. A higher rotation
rate will lead to a larger separation. But the characteristic of
the parallel separation seems to be not in good agreement with that
of NGC 1856.

\textbf{If there exist a large \dov{} and a rapid rotation in stars, 
the elements produced in the core can be easily brought to the surface
of the stars. For the model with \dov{} $=0.6$ and $\Omega_{i}=0.5$
in young clusters, the surface nitrogen abundance at the end of 
MS is about $1.5$ times as large as the initial abundance. 
But for the models with \dov{} $=0.2$ and $\Omega_{i}=0.5$ and 
the models with \dov{} $\leq0.6$ without rotation, the surface 
nitrogen abundance is almost invariable in MS phase. 
If there is a nitrogen spread in MSTO stars that would reflect 
the coexistence of rapid rotation and large \dov{}. 
Due to the mixing effect of deep convective envelope, 
the surface nitrogen abundance in RGB is 
about $1.5-3$ times as large as the initial abundance. 
But for the cluster with an age of $550$ Myr, the envelopes of 
RGB stars with \dov{} $=0.7$ and $\Omega_{i}=0$ are radiative.
Their nitrogen abundance cannot be enhanced. For RGB stars 
with an age $> 4$ Gyr, the surface nitrogen abundance of models
with \dov{} $=0.6$ is about $1.4$ times as large as that of models 
with \dov{} $=0$ in non-rotation case, but is about $1.7$ times
in rotation case.
}

\subsubsection{Variable overshooting scenario}

In Figure \ref{fig10}, we compare overshooting models with the same
age but with different \dov{} to the data of NGC 1856. Panel $a$ of
Figure \ref{fig10} shows that the eMSTOs of NGC 1856 can be explained
well by the effects of \dov{} between 0 and 0.7. The separation between
the isochrone with \dov{} $=0.7$ and that with \dov{} $=0$ is about
0.1 mag at $m_{U}$ $\simeq20.0$ and is around 0.014 mag
at $m_{U}$ $\simeq21.5$.

In order to understand whether or not overshooting models can produce
the eMSTOs and the split MSs of NGC 1856, we performed a stellar population
synthesis by way of Monte Carlo simulations following the initial mass
function of \cite{salp55}. \cite{milo15} found that
about $70\%$ and $30\%$ of stars belong to the blue- and red-MS in NGC 1856,
respectively. We assumed $10$\% of population with \dov{} $=0.7$,
$20$\% of population with \dov{} $=0.6$, $40$\% of population with \dov{} $=0.4$,
$20$\% of population with \dov{} $=0.2$, and $10$\% of population with \dov{} $=0$.
In the synthesized population, we included observational errors taken
to be a Gaussian distribution with a standard deviation of 0.03
in magnitude and color. Panel $b$ of Figure \ref{fig10} shows that
the eMSTO of NGC 1856 is reproduced well by the simulated coeval population.

Moreover, panel $a$ of Figure \ref{fig10} shows that the bottom of the RGB
of our variable \dov{} models is located at U-V $\sim1.2$ and $m_{U}$
$\sim20.7$. Both the observational and the simulated results show that
there are many stars. These stars cannot appear in the simulations
of eSFH models and rapid rotation models.

\textbf{The intermediate-age star clusters have a relatively tight 
SGB \citep{lic14, lic16a}.} The subgiants of NGC 1856 seem to
be not reproduced well by the simulation.
The larger the \dov{}, the shorter the timescale of subgiants.
Moreover, the number of stars with \dov{} $=0.6$ and $0.7$ are
less than that of stars with \dov{} $=0.4$.
Thus, it is hard for the subgiants with \dov{} $=0.6$ and $0.7$
to occur in the CMD. The number of stars with \dov{} $=0$ are
much less than that of stars with \dov{} $=0.4$. Therefore,
the number of subgiants obtained by the simulation are also rare.
In order to understand why there are many stars with U-V between
about 0.6 and 1.2 in the CMD of NGC 1856, we calculated
a binary-star stellar population following \cite{yang11} by
using the Hurley rapid binary evolution codes \citep{hurl02}.
The calculation shows that the effects of binary stars are not
enough to explain the eMSTOs of NGC 1856. However, the calculation
clearly shows that there are many interacting binary stars
with core-helium burning (CHeB) whose luminosity and the effective
temperature are very close to that of subgiants (see
Figure \ref{fig11}). The subgiant branch of NGC 1856 could be
polluted by the CHeB interacting binary stars.
Many of stars of NGC 1856 with U-V between about $0.6$
and $1.2$ and with $m_{U}$ around $18.5$ mag
might be the CHeB interacting binaries. Moreover,
the CHeB interacting binaries could make us misunderstand
that the ``subgiants'' of NGC 1856 mainly concentrate towards
the end of the blue-MS expected from the MSTOs of the cluster.

When both the effects of a rapid rotation and that of different OVCCs
exist in the star cluster NGC 1856, Figure \ref{fig12} clearly shows
that the effects of the rapid rotation of $0.5$ $\Omega_{cr}$ on the eMSTOs
are negligible compared with the effects of the different \dov{}.
The eMSTOs of the isochrones mainly result from the effects of the OVCC.
The separation between the isochrone with \dov{} $=0.7$ and $\Omega_{i}=0$
and that with \dov{} $=0.2$ and $\Omega_{i}=0.5$ is about
$\mathbf{0.09}$ mag at $m_{U}$ $\simeq20.0$ and is around $\mathbf{0.028}$ mag
at $m_{U}$ $\simeq21.5$. The centrifugal effect increases
with an increase in rotation rate. A higher rotation rate
will lead to the more significant split at $m_{U}$ $\sim21$,
which cannot be mimicked by the effects of OVCC.
Thus, the determinations of rotation rates of stars and the
split of MS of young clusters at $m_{U}$ $\lesssim21$ aid in
determining the effects of rotation.

\textbf{The split MS of NGC 1856 was not reproduced by OVCC models.
This could not rule out OVCC scenario because variable \dov{} and 
rapid rotation could coexist in a star cluster. 
Moreover, the blue MS could be affected by interacting binaries
(see Figure \ref{fig11}). }

\subsubsection{A prediction of OVCC models}
\textbf{There is a pulsation constant 
\begin{equation}
 Q=\mathit{\Pi}(\frac{M}{M_{\odot}})^{1/2}(\frac{R}{R_{\odot}})^{-3/2}
\end{equation}
for pulsating stars, where $\mathit{\Pi}$ is a period of pulsating stars,
or a large frequency separation
\begin{equation}
 \Delta\nu=134.9\times(\frac{M}{M_{\odot}})^{1/2}(\frac{R}{R_{\odot}})^{-3/2} \mu\mathrm{Hz}
\end{equation}
for stars with solar-like oscillations. In a star cluster, 
the radii of MSTO stars with a large \dov{} are much larger
than those of MSTO stars with a small \dov{}. Thus, 
the periods of the former could be about $2-4$ times as long
as those of the latter. For example, at the end of the 
central H-burning phase of a star with $M=3.0$ \dsm{},
the period of the model with \dov{} $=0.6$ is about $4.4$
times as long as that of the model with \dov{} $=0$. 
In other words, if the variable \dov{}
exists in a star cluster, long-periodic pulsating/variable stars 
(in bright MSTO) and short-periodic pulsating stars
(in faint MSTO) would be observed in the star cluster. 
This case cannot occur in rotation scenario and eSFH scenario.
For example, a difference of $300$ Myr in age can only lead to 
a change of about $10\%$, depending on the age, in frequencies 
or periods of pulsation of MSTO stars. Gravity darkening 
caused by the angle of view cannot result in any change 
in the frequencies or periods. Thus, asteroseismical observation
for the young- and intermediate-age star clusters would directly
rule out or confirm OVCC scenario, and show what plays 
a dominant role in the origin of the eMSTO.
}

\section{DISCUSSION AND SUMMARY}

\begin{figure}
\centering
\includegraphics[scale=0.7, angle=-90]{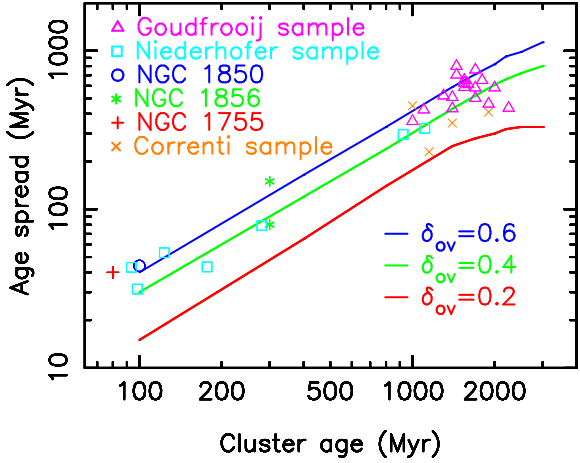}
\caption{Comparison between the equivalent age spreads of MSTOs of star clusters
caused by the OVCC and those observed in different clusters.
The large and the small spreads of NGC 1856 are given by \cite{milo15}
and \cite{corr15}, respectively. The age spreads of Correnti sample
are listed in \cite{piat16}.}
\label{fig5}
\end{figure}

\begin{figure}
\centering
\includegraphics[scale=0.34, angle=-90]{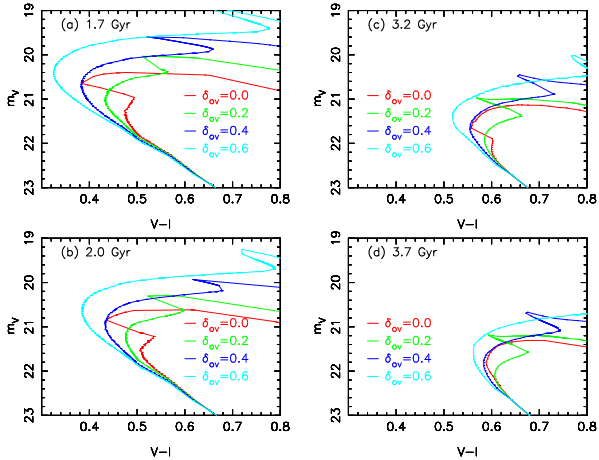}
\caption{The CMD of isochrones with different \dov{} at different ages.}
\label{fig6}
\end{figure}

\begin{figure}
\centering
\includegraphics[scale=0.36, angle=-90]{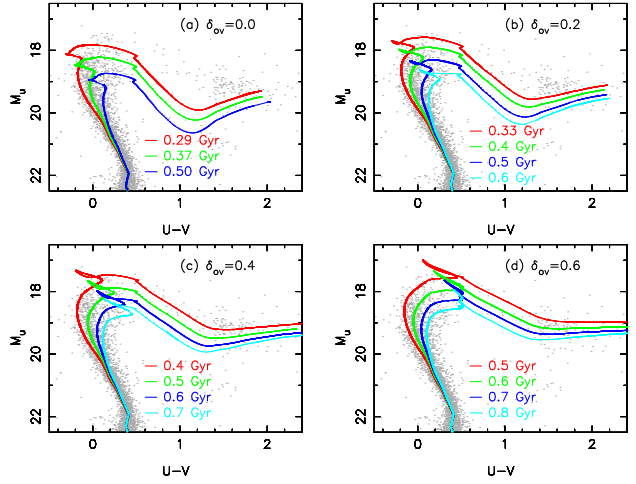}
\caption{Comparison between the data of NGC 1856 and isochrones with
different \dov{} and ages. The grey dots represent the observed stars
of NGC 1856 \citep{milo15}. A distance modulus of $18.6$ is adopted,
and the value of $0.12$ is adopted for $E(U-V)$ in the calculation.}
\label{fig7}
\end{figure}

\begin{figure}
\centering
\includegraphics[scale=0.36, angle=-90]{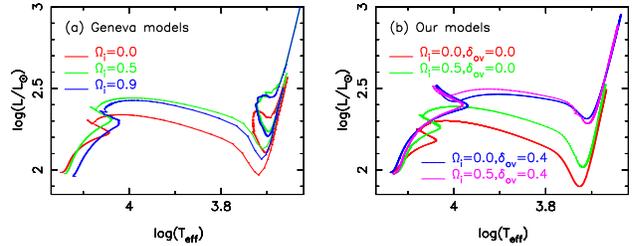}
\caption{Evolutionary tracks of models with $M=3.0$ \dsm{}. The value of
$Z_{i}$ is 0.006 for Geneva models but is 0.008 for our models.}
\label{fig8}
\end{figure}

\begin{figure}
\centering
\includegraphics[scale=0.5, angle=-90]{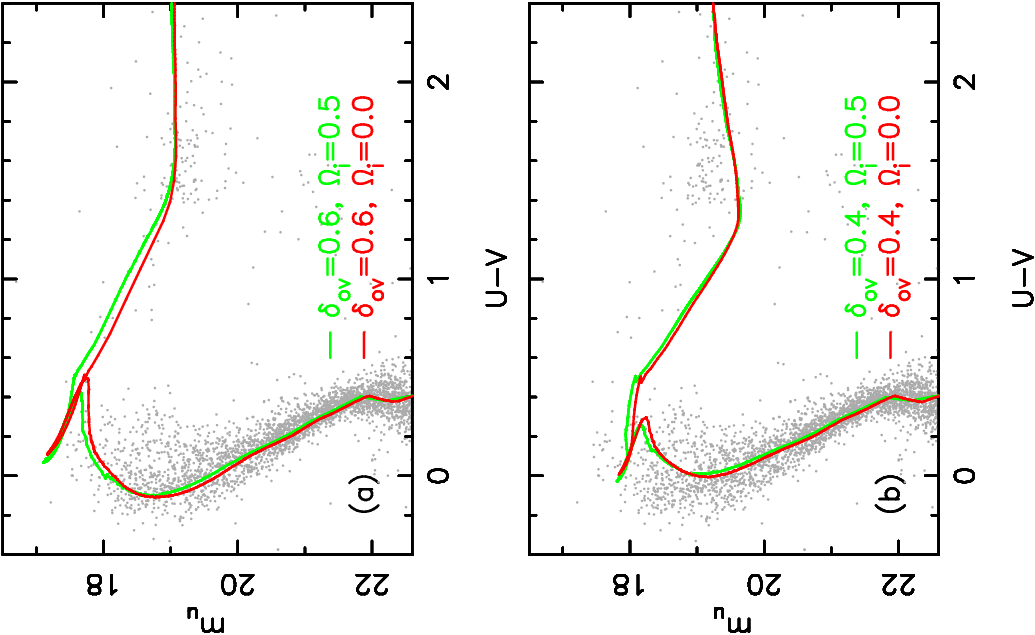}
\includegraphics[scale=0.5, angle=-90]{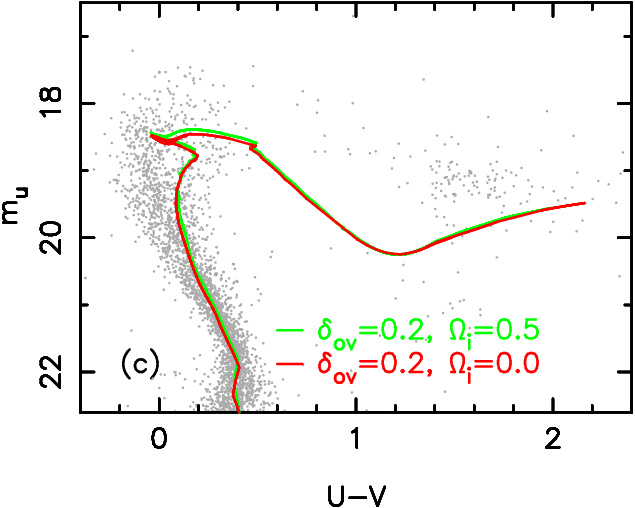}
\caption{The CMDs of NGC 1856 and rotating and non-rotating isochrones
with an age of $550$ Myr. The values of \dov{} for the isochrones in
panels $a$, $b$, and $c$ are 0.6, 0.4, and 0.2, respectively.}
\label{fig9}
\end{figure}

\begin{figure}
\centering
\includegraphics[scale=0.6, angle=-90]{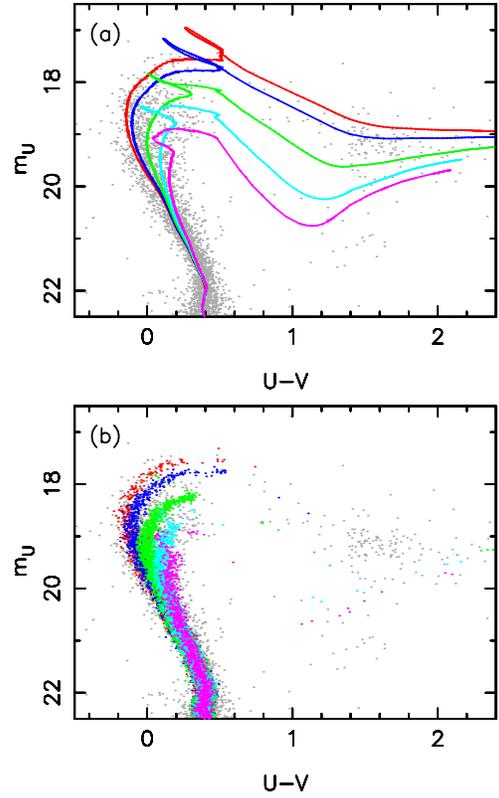}
\caption{Panel $a$ shows the comparison between the data of NGC 1856
and the isochrones with an age of $550$ Myr but with different \dov{}.
The red line represents the isochrone with \dov{} $=0.7$, the blue line
\dov{} $=0.6$, the green line \dov{} $=0.4$, the cyan line
\dov{} $=0.2$, the magenta line \dov{} $=0$.
Panel $b$ shows the comparison between the data of NGC 1856 and the
simulated populations with an age of $550$ Myr but with different \dov{}.}
\label{fig10}
\end{figure}

\begin{figure}
\centering
\includegraphics[scale=0.6, angle=-90]{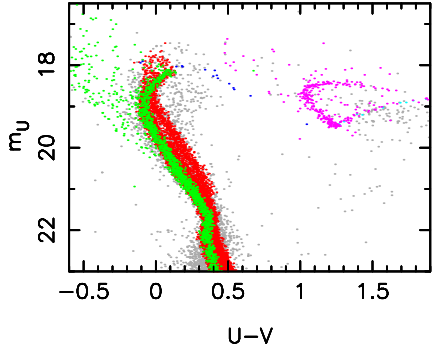}
\caption{Comparison between the data of NGC 1856 and simulated binary
populations with an age of 450 Myr. Red dots represent non-interacting
binary systems in MS. Green dots show interacting binary systems in MS,
blue dots subgiants, cyan dots RGB stars, magenta dots core-helium
burning binary stars. A distance modulus of $18.5$ is adopted,
and the value of $0.08$ is adopted for $E(U-V)$ in the calculation.}
\label{fig11}
\end{figure}

\begin{figure}
\centering
\includegraphics[scale=0.5, angle=-90]{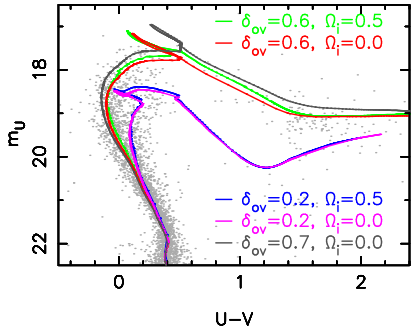}
\caption{The CMD of NGC 1856 and isochrones with an age of $550$ Myr but with
different \dov{} and $\Omega_{i}$.}
\label{fig12}
\end{figure}

Figures \ref{fig5} and \ref{fig10} show that the OVCC can lead to
the observed eMSTOs. The variable OVCC is a good potential
mechanism for explaining the eMSTOs of star clusters. However,
it is dependent on the fraction of stars with a large \dov{} that
whether or not the OVCC is responsible for the observed eMSTOs of star
clusters. Asteroseismology only tells us that the value of \dov{}
can be different for different stars with an approximatelly equal mass.
It has not revealed the fraction of stars with a large \dov{}.
Why there are different \dov{} for stars is another open question
for asteroseismology and stellar physicists.

The OVCC is more efficient than rotation at bringing H-rich material
into the H-burning core. The effects of the OVCC with \dov{} $=0.4$
on the MSTO of star clusters is hard to be mimicked by the effects
of a rapid rotation. In order to reproduce the stars of NGC 1856 at
the end of central H-burning phase, an overshooting of around
0.4 $H_{p}$ is required for our models. When both the OVCC and
a rapid ratation exist in stars, the MSTO is mainly affected
by the effects of the OVCC. However, the centrifugal effect
of the rapid rotation playing a role in the MS stars whose
envelope is radiative leads to the fact that the split
of MS is more significant.

The points C and D of the rotating isochrones with
$\Omega_{i}=0.5$ can be matched by those of non-rotating
isochrones with the same \dov{} but with different ages. The equivalent
age spread obtained by comparing points C and D of a rotating isochrone
to those of non-rotating isochrone with the same \dov{} is
twice more than that obtained through comparing points A and B
of the rotating isochrone to those of non-rotating isochrone.
\cite{nied15b} might overestimate the equivalent age spreads
caused by rotation. Moreover, the higher the efficiency of
the rotational mixing, the more the H-rich material brought
into the H-burning core, the larger the luminosity of the
rotating models at the end of central H-burning phase,
which can lead to an increase in the equivalent age spread
caused by rotation. These may be the reason why the equivalent
age spread of \cite{nied15b} rotating models can match that
of NGC 1856 but that of our rotating models cannot. However,
a high efficiently rotational mixing in the radiative
region of stars can hinder stellar expansion and
lead to the fact that the evolutionary tracks of the
stars move towards the upper left of the Hertzsprung-Russell
diagram. Thus, the effects of a large \dov{} are different
from that of rotation.

The equivalent age spread between the isochrone with \dov{}
$=0.4$ and that with \dov{} $=0$ in panel $a$ of Figure \ref{fig10}
is $170$ Myr, and the spread between the isochrone with \dov{}
$=0.6$ and that with \dov{} $=0.2$ is $180$ Myr, which are
close to the spread of 150 Myr determined by \cite{milo15}.
The equivalent age spread between the isochrone with \dov{}
$=0.6$ and that with \dov{} $=0$ is $230$ Myr, which is larger
than that determined by \cite{milo15}. However, Figure \ref{fig10}
shows that the width between the MSTO of the isochrone with \dov{}
$=0.6$ and that with \dov{} $=0$ does not exceed that of
NGC 1856.

Due to the fact that the lifetimes of the SGB of the stars
with a large \dov{} are much shorter than those of the stars
without overshooting or with a small \dov{}, the population
of SGB is not only dependent on the number of stars but also
dependent on \dov{}. Moreover, the SGB of young star clusters
could be polluted by the CHeB interacting binaries.
Thus, the SGB of young star clusters cannot be used
to study the origin of the eMSTOs.

\textbf{The effects of OVCC on the MSTO of young- and 
intermediate-age star clusters
are similar to that of rotation. The equivalent age spread
caused by rotation has a peak at age $\sim1.7$ Gyr \citep{yang13}, 
which is dependent on the efficiency of rotational mixing. 
The observationally inferred age spread also has a peak 
at age $\sim1.7$ Gyr \citep{nied16, lic16b}. This characteristic 
does not occur in OVCC models. The eMSTO caused by the OVCC 
is still present in star clusters with age $> 2.0$ Gyr. The 
eMSTO is hard to be exactly described by the equivalent
age spreads relative to the same age isochrone with \dov{} $=0$,
especially for star clusters with age $\gtrsim 3.0$ Gyr. 
Our models show that the width of the eMSTO caused by a 
\dov{} of $0.2$ or $0.4$ at age $\sim2.5$ Gyr is about $0.5$
times as wide as that at age $\sim1.4$ Gyr.
The width of the eMSTO caused by the OVCC has a peak in color V-I 
of CMD of star clusters at age $\sim 1.4$ Gyr, and then decreases
with an increase or decrease in age, which cannot be reflected 
by the equivalent age spreads. In rotation models of \cite{yang13},
the eMSTO can occurs in star clusters with age larger than $2.0$ Gyr 
because rotational mixing exists in stars like the Sun \citep{yang07, yang16a}.
Thus the eMSTO of star clusters with age $>2.0$ Gyr cannot be used to
differentiate OVCC models and rotation models. If eMSTO does not 
exist in star clusters with ages between $2.0$ and $2.5$ Gyr 
that is hard to be explained by OVCC scenario. Perhaps, that 
is due to the fraction of stars with a large \dov{} being small 
in the star clusters.}

\textbf{
At the end of the central H-burning phase of rapid rotation models,
the surface nitrogen abundance can be enriched by $25-50\%$ 
in young star clusters for a \dov{} between 0.4 and 0.6. 
This does not happen in the rotating models with \dov{} $= 0.2$ 
and the non-rotating models. Moreover, the pulsating periods of bright
MSTO stars would be longer than those of faint MSTO stars in OVCC scenario
because the former has a low mean density, which also does not
occur in rotation scenario and eSFH scenario. These characteristics
could aid us in differentiating OVCC scenario and rotation scenario.}

Asteroseismology has shown that the value of overshooting parameter of
the convective core is variable for stars with an approximatelly equal
mass. We calculated the effects of the OVCC on the MSTO
of young- and intermediate-age star clusters, and found that
the observed eMSTOs of young- and intermediate-age star clusters
can be explained well by the effects of the OVCC with different \dov{}.
The equivalent age spread caused by the OVCC increases with
an increase in age of star clusters and an increase in \dov{}.
Due to the fact that the convective core of stars can shrink
during the late stage of MS, the equivalent age spread cannot
increase continuously with an increase in age of star clusters.
The eMSTOs of NGC 1856 can be reproduced well by the coeval
populations with different \dov{}. The effects of a rotation
with $\Omega_{i}=0.5$ and the rotational mixing efficiency
for the Sun on the MSTO of a cluster with an age of $550$ Myr
are negligible compared with the effects of the OVCC.
However, the presence of the rotation can result in the fact
that the split of MS of the cluster is more significant.
The eMSTOs of star clusters may be related to the effects 
of the OVCC with different \dov{}. \textbf{But keep in mind 
that a single value of $\delta_{\rm ov}$ cannot produce 
any eMSTO in the CMD of clusters. Asteroseismical observation 
would confirm or rule out OVCC scenario. }

\acknowledgments
We thank the anonymous referee for helpful comments that
helped the authors improve this work, A. P. Milone for
providing the data of NGC 1856, and F. Niederhofer for
the data of Niederhofer sample, as well as the support
from the NSFC 11273012.

\clearpage

\end{document}